\documentclass[twocolumn,showpacs,preprintnumbers,amsmath,amssymb]{revtex4}

\usepackage{graphicx}
\usepackage{dcolumn}
\usepackage{bm}
\usepackage{here}

\begin{document}

\preprint{Revtex4}

\title{
Localization Properties of Electronic States in Polaron Model of poly(dG)-poly(dC) and
poly(dA)-poly(dT) DNA polymers 
}

\author{Hiroaki Yamada}
 \email{hyamada@uranus.dti.ne.jp.}
\affiliation{%
Aoyama 5-7-14-205,
Niigata 950-2002,
Japan
}%
\author{Eugen B. Starikov}
 \email{starikow@chemie.fu-berlin.de}
\affiliation{
Karolinska Institute, Center for Structural Biochemistry NOVUM,
S - 14157 Huddinge, Sweden
}%
\author{Dirk Hennig}
 \email{hennigd@12move.de}
\affiliation{
Freie Universit$\ddot{a}$t Berlin, Fachbereich Physik,
Institut f$\ddot{u}$r Theoretische Physik, Arnimallee 14, 14195 Berlin,
Germany
}%
\author{Juan F.R. Archilla}
 \email{archilla@us.es}
\affiliation{Group of Nonlinear Physics, Departamento de F$\acute{s}$ica Aplicada I, ETSI Inform$\acute{a}$tica,
University of Sevilla, Avda Reina Mercedes s/n, 41012 - Sevilla, Spain
}%

\date{\today}

\begin{abstract}
We numerically investigate localization
properties of electronic states in a
 static model of poly(dG)-poly(dC) and poly(dA)-poly(dT) DNA
polymers with realistic parameters
obtained by quantum-chemical calculation. The randomness in the
on-site energies caused by the electron-phonon coupling are
completely correlated to the off-diagonal parts.
In the single electron model, the effect of the hydrogen-bond
 stretchings,   the twist angles
 between the base pairs  and the finite system size effects
 on  the energy dependence of the localization length and on the Lyapunov
 exponent are given.
 The localization length is reduced by the
influence of the fluctuations
in the hydrogen bond stretchings.
It is also shown that the helical twist angle affects the
localization length in the poly(dG)-poly(dC) DNA polymer
 more strongly than in  the poly(dA)-poly(dT) one.
 Furthermore, we show
resonance structures
 in the energy dependence of the localization
length when the system size is relatively small.
\end{abstract}

\maketitle

\section{Introduction}

The recent development of the nanoscale fabrication let us expect
the utilization of the DNA wire as a molecular device
\cite{lewis03,porath04} and the realization of DNA computing
\cite{paun98}. DNA is believed to form an effectively
one-dimensional molecular wire, which is highly promising for
diverse applications. Actually,
 the modern development of physico-chemical experimental techniques
enables us to measure directly
 DNA electrical transport phenomena even
in single molecules \cite{porath04,tran00}. Moreover, several
groups have recently performed numerical investigations of
localization properties of DNA electronic states based on
realistic DNA sequences
\cite{carpena02,roche03,yamada04a}.

However, DNA transport properties still remain a controversial
topic, mainly due to tremendous difficulties in setting up the
proper experimental environment and the
complexity of the molecule itself. Specifically, a distinctive
feature of biological polymers is the
complicated composition
 of their elementary subunits, and an apparent ability of their
structures to support long-living nonlinear excitations. Although
a number of theoretical explanations for DNA charge
transfer/transport phenomena have been suggested on the basis of
standard solid-state-physical approaches, like polarons, solitons,
electrons or holes
\cite{lewis03,hennig02,chang03,conwell00,campbell04,hermon98,lebard03,iguchi03,yamada04},
the situation is still far from working out a unique,
non-contradictory theoretical scheme.

In their polaron-like model, Hennig and coworkers studied electron
breather propagation along DNA homopolynucleotide duplexes, i.e.
in both poly(dG)-poly(dC) and poly(dA)-poly(dT) DNA polymers
\cite{hennig02}, and, for this purpose, estimated
electron-vibration coupling strength in DNA using semiempirical
quantum-chemistry \cite{hennig04,starikov02}. Chang \textit{et
al.} have also considered
a possible mechanism to explain the phenomena of DNA charge
transfer \cite{chang03}. The charge coupling with DNA structural
deformations  can create a polaron and thus promote a
localized state \cite{chang03,conwell00}. As a result, the moving
electron breather may
 contribute to highly efficient long-range
conductivity. Recent experiments seem to support the polaron
mechanism for the electronic transport in DNA polymers
\cite{yoo01}.

In the present paper, we investigate localization properties of
electronic states in a
stochastic bond vibration model of poly(dG)-poly(dC) and
poly(dA)-poly(dT) DNA polymers by adopting the model by Hennig
{\it et al.} \cite{hennig02}. Here we assume that the disorder is
caused by DNA vibrational modes, and via electron-vibrational
coupling it influences the charge transfer/transport along DNA
duplexes. Moreover, we discuss the difference in charge
localization properties between poly(dG)-poly(dC) and
poly(dA)-poly(dT) DNA polymers, as well as the peculiarities of
the mixed (AT/GC) model.

The outline of the present paper is as follows. In the next section we
introduce the DNA model to investigate subsequently
the electronic states and their coupling to DNA vibrational modes.
In the Sec. 3 we present numerical results concerning the
influence of changes of the hydrogen-bond stretchings and twist
angles as well as the effects of finite system size on the
localization properties. The last section contains a
summary and discussion.

\section{Model}

This section reformulates the model Hamiltonian for DNA duplexes,
as introduced by Hennig {\it et al.} \cite{hennig02} into a
one-electron tight-binding adiabatic Hamiltonian with structural
disorder, taking into account a
nearest-neighbor electron hopping term. The latter model is actually
akin to that proposed earlier by Dunlap-Kundu-Philips (DKP). Specifically,
the general DKP model contains disorder induced by the proper vibrational
modes, which shows up in both
 the diagonal and the off-diagonal parts
 of the electron Hamiltonian \cite{dunlap89}. 
Generally, in the
DKP model the transport can arise from a set of measure-zero
unscattered states at a particular energy derived from the resonant transmission formula 
\cite{dunlap89,varga98}, and the number of the unscattered states can be estimated 
by the resonance width in the system size.

The Hamiltonian
for the electronic part in our DNA model is given by
 \begin{eqnarray}
H_{el} & = & \sum_{n}E_{n}C_{n}^{\dagger}C_{n}  \nonumber \\
&-& \sum_{n}V_{nn+1}(C_{n}^{\dagger}C_{n+1}+C_{n+1}C_{n}^{\dagger}),
\end{eqnarray}
 where $C_{n}$ and $C_{n}^{\dagger}$ are creation and annihilation
operators of an electron at the site $n$. The on-site energies $E_{n}$
are represented as
\begin{eqnarray}
E_{n} & = & E_{0}+kr_{n},
\end{eqnarray}
 where $E_{0}$ is a constant and $r_{n}$ denotes the fluctuation
caused by the coupling with the transversal Watson-Crick H-bonding stretching
vibration.

The transfer integral $V_{nn+1}$ depends on the three-dimensional
stacking-distance $d_{nn+1}$ between adjacent bases labeled by $n$ and
$n+1$, along each strand and is given as,
\begin{eqnarray}
V_{nn+1} & = & V_{0}(1-\alpha d_{nn+1})\,.
\end{eqnarray}
The parameters $k$ and $\alpha$, which describe the strength of
the interaction between the electronic and spatial variable, have
been previously calculated through quantum-chemical
 methods.
Radial displacements bring about also a variation of the distances
between neighboring bases along each strand $d_{nn+1}$. The
 first--order Taylor expansion around the equilibrium positions is given by
\begin{eqnarray}
d_{nn+1} & = & \frac{R_{0}}{\ell_{0}}(1-\cos\theta_{0})(r_{n}+r_{n+1}).
\end{eqnarray}
 $R_{0}$ represents the equilibrium radius of the helix, $\theta_{0}$
is the equilibrium twist angle between base pairs, and $\ell_{0}$
the equilibrium distance between bases along a strand given by
\begin{eqnarray}
\ell_{0} & = & (a^{2}+4R_{0}^{2}\sin^{2}(\theta_{0}/2))^{1/2},
\end{eqnarray}
 with $a$ being the distance between neighboring base pairs in the
direction of the helix axis. A sketch of the geometrical
parameters $R_{0},\ell_{0},\theta_{0},r_{n+1}$ and $d_{nn+1}$ is
given in Fig.1.

\begin{figure}[h]
\includegraphics[scale=.65]{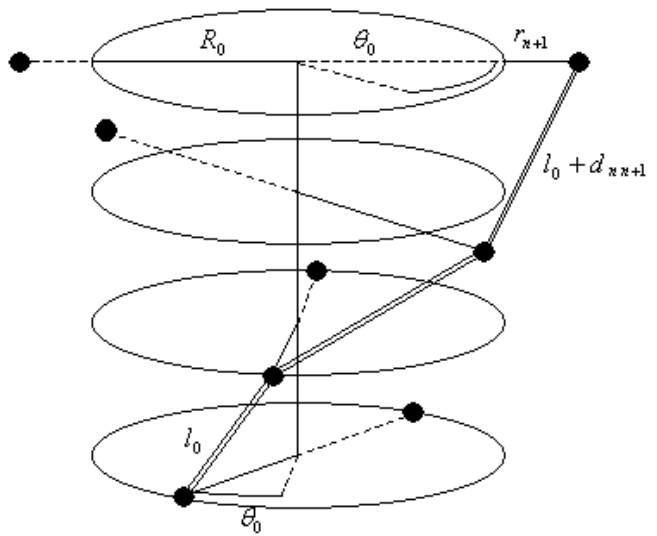}
 \caption{
Sketch of the structure of the DNA model. The bases are represented
by bullets and the geometrical parameters $R_0, \ell_0, \theta_0, r_{n+1}$
 and $d_{nn+1}$
are indicated.
}
\end{figure}

We adopt realistic values of the parameters obtained from
the semi-empirical
quantum-chemical calculations; $k=0.778917 [eV$\AA$^{-1}]$,
$\alpha=0.053835 [$\AA$^{-1}]$ for the coupling parameters of
the poly(dA)-poly(dT) DNA polymer, $k=-0.090325 [eV$\AA$^{-1}]$,
$\alpha=0.383333 [$\AA$^{-1}]$ for
 the ones of
the poly(dA)-poly(dT) DNA polymer. As to
 the other typical parameters for
DNA molecules, we use: $E_{0}=0.1 [eV],V_{0}=0.1 [eV]$, $a=3.4
[$\AA$], R_{0}=10 [$\AA$]$ and $\theta_{0}=36^{\circ}$.

Further, we consider $\{ r_{n}\}$ as
independent random variables generated by uniform distribution
with width ($r_{n}\in[-W,W]$). Accordingly, fluctuations
 in both the
 on-site energies and the off-diagonal parts in the Hamiltonian (1)
are mutually correlated because they are generated by the same
random sequence $r_{n}$. (See Fig.2(c).) The typical value for $W$
is $W=0.1 [$\AA$]$, which approximately corresponds
 to the variance in the hydrogen bond lengths in
Watson-Crick base pairs, as seen in X-ray diffraction experiments
\cite{xray}.

Figures 2(a) and (b) show a typical fluctuation pattern
of $E_{n}$ and $V_{nn+1}$ for $W=0.1$ in poly(dG)-poly(dC) and
poly(dA)-poly(dT) DNA polymers. Although the fluctuation of the
on-site energy is almost of
the same order in both poly(dG)-poly(dC) and poly(dA)-poly(dT) DNA
polymers, the fluctuation of the transfer integral $V_{nn+1}$ in
the poly(dG)-poly(dC) DNA polymers is much larger than that of the
poly(dA)-poly(dT) DNA polymers. This property ought to reflect the
difference in the electron localization nature between
poly(dG)-poly(dC) and poly(dA)-poly(dT) DNA polymers. We can see
the correlation between the sequences, $E_{n}$, $V_{nn+1}$, in the
parametric plots in Fig.2(c).

In addition to the DNA homopolymer duplexes, we investigate the
localization nature
of the mixed sequence consisting of two types
of the Watson-Crick pairs. Then, as a zero-order approximation,
the electron-phonon coupling parameters for the mixed GC/AT stacks
are taken here to be equal to the values obtained for
poly(dG)-poly(dC) and poly(dA)-poly(dT) DNA polymers.

\begin{figure}[!h]
\includegraphics[scale=.6]{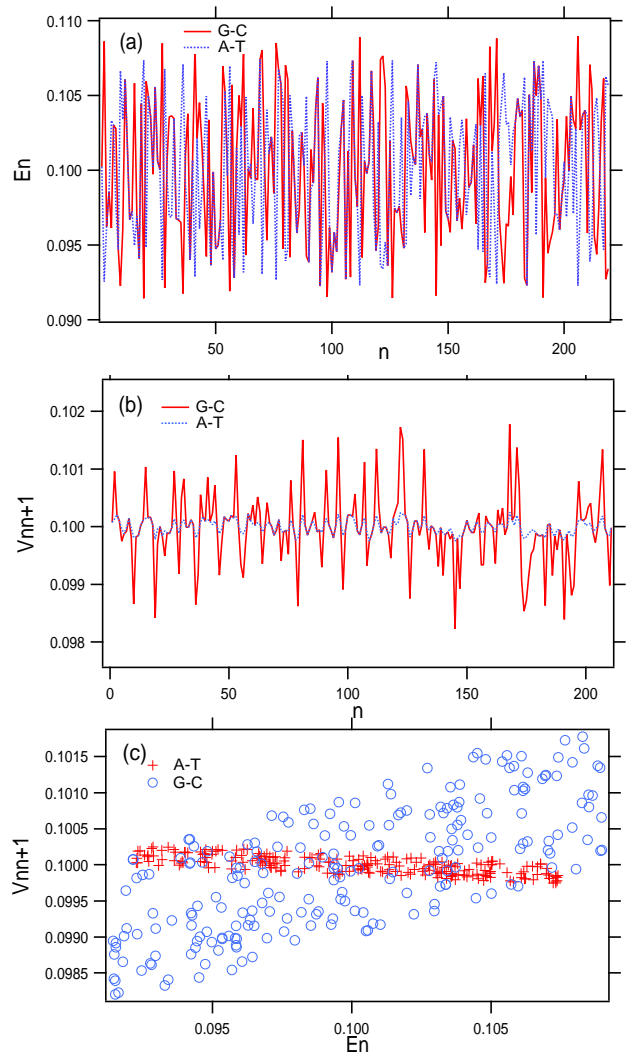}
 \caption{(a)The on-site energy $E_n[eV]$ and (b)transfer integral
 $V_{nn+1}[eV]$ as a function of the base pair
site $n$. The parametric plot $E_n$ versus $V_{nn+1}$ is shown in (c).
$W=0.1$ and the other parameters are given in text. 
The unit of the energy and the spatial length are $[eV]$ and 
the number of nucleotide base pair $[bp]$, respectively, throughout the present paper. 
}
\end{figure}

\section{Numerical Results}
The Schr$\ddot{o}$dinger equation $H_{el}|\Phi>=E|\Phi> $ is written in the transfer matrix form,
\begin{eqnarray}  
 \left(
\begin{array}{c}
a_{n+1} \\
a_{n} \\ 
\end{array}
\right)
=
\left(
\begin{array}{cc}
\frac{E-E_{n} }{V_{nn+1}} & -\frac{V_{nn-1}}{V_{nn+1}}  \\
 1 & 0 
\end{array}
\right)
\left(
\begin{array}{c}
a_{n} \\
a_{n-1} \\ 
\end{array}
\right),
\end{eqnarray}  
\noindent
 where $a_{n}$ is the
 amplitude of the electronic wavefunction
 $|\Phi\rangle =\sum_n a_n |n\rangle $
 at  the base pair site $n$.
 We use the localization length $\xi$ and/or 
Lyapunov exponent $\gamma $ calculated by the mapping (6) 
in order to characterize the exponential 
localization of the wave function. 
   Originally the Lyapunov exponent (inverse localization length) is
defined in the thermodynamic limit ($N\rightarrow\infty$),
however, in the present paper we use the
following definition for the Lyapunov exponents  of the electronic wave function for a
large system size $N$ \cite{crisanti93,yamada01}.
\begin{eqnarray}
\gamma(E,N)=\xi^{-1}(E,N)=\frac{\ln(|a_{N}|^{2}+|a_{N-1}|^{2})}{2N}.
\end{eqnarray}
 We use appropriate initial conditions
$a_{0}=a_{1}=1$, and  for large $N(>>\xi)$ the localization length and
Lyapunov exponent are independent of the boundary condition. 
The system size dependence of the localization
length is investigated in Subsect.3.3 in the relation with
 the resonance energy.
  The energy-dependent transmission coefficient $T(E,N)$ of the system between 
metallic electrodes is given as $T(E,N)=\exp (-2\gamma N)$ and is related to 
Landauer resistance via $\varrho =(1-T)/T$ 
in unit of the quantum resistance $h/2e^2( \sim 13[k\Omega m])$ \cite{lifshits88}.

\subsection{Fluctuation effect of stretching}
Figure 3(a) shows
  the dependence of the localization length $\xi$ on the energy
 for the cases of $W=0.1,0.2,0.4$ in the poly(dA)-poly(dT) DNA polymer. 
  For the computation
of the energy dependence of the localization length we used a
energy subdivision of 800 points in the energy range.
  The extent of the localization depends on the Fermi energy of the electroads when
we measure  the transmission coefficient or conductance.
It is apparent
 that the localization length is
reduced as a result of the
increase of the fluctuation strength $W$ of the hydrogen bond
stretching. The small peaks for the resonance energy are
 suppressed in $W\geq0.2$ due to the large fluctuation in
the $r_{n}$.

Figure 3(b) and (c) show the localization length as a function of
the energy
in the case of poly(dG)-poly(dC) and the mixed DNA polymers,
respectively. In Fig.3(b) the global behavior is almost the same
as that illustrated in Fig.3(a)
except for the shape of the energy dependence. The localization
length in the
poly(dA)-poly(dT) DNA polymer is globally larger than the
 one in poly(dG)-poly(dC)
DNA polymer.
 The reason is simply because the
fluctuation strength of $V_{nn+1}$ in poly(dG)-poly(dC) DNA
polymers is larger than the one in poly(dA)-poly(dT) DNA
polymers as seen in Fig.2.

In the mixed case, the localization length in the mixed sequence
shows the intermediate behavior between the poly(dG)-poly(dC) and
poly(dA)-poly(dT) DNA
polymers.
The resonance peak structure in the energy dependence of the localization length 
is not as pronounced as in
the case of poly(dA)-poly(dT) and poly(dG)-poly(dC) DNA polymers.

As a result, more disorder in the H-bond stretching (higher W
values) renders
the peak structure in
the electronic states to disappear and makes the latter
more localized. 
  The finite system effect will be discussed in Subsect.3.3.

\begin{figure}[!h]
\includegraphics[scale=.55]{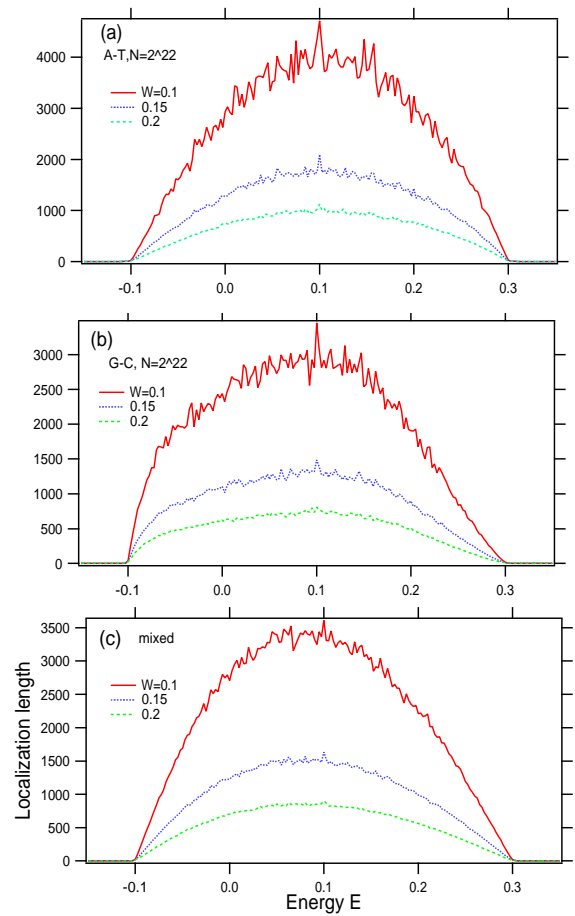}
 \caption{The Localization length as a function of
 energy for several values of
$W(=0.1,0.15,0.2)$ in the (a) poly(dA)-poly(dT) DNA polymers, (b)
poly(dG)-poly(dC)  DNA polymers and (c)  mixed DNA polymers. The
system size is
$N=2^{22}$ and $\theta_0=36^\circ$. }
\end{figure}

\subsection{Twist angle effect}
In this subsection, we investigate how the localization length
depends on the helical twist angle $\theta_{0}$ of the
DNA double helix. It is known that the allowed  values of the
angle range from ca. $27^{\circ}$ till $45^{\circ}$.

Figure 4 shows the energy dependence of the localization length
for some twist angles in the DNA models.
We conclude, that in poly(dA)-poly(dT)
polymers the change of the angle does virtually not affect the
localization length, however, in the poly(dG)-poly(dC) polymer it
does have an impact.
The difference is caused by the $\theta_{0}$ dependence of the
fluctuation of the transfer energy $V_{nn+1}$ which is smaller in
the case of the poly(dG)-poly(dC) polymer than in its
poly(dA)-poly(dT) counterpart as seen in Fig.2(b).

It follows that when the twist angle increases in the
poly(dG)-poly(dC) DNA polymer the localization length becomes
smaller (larger) in the low (high) energy region.

\begin{figure}[!h]
\includegraphics[scale=.55]{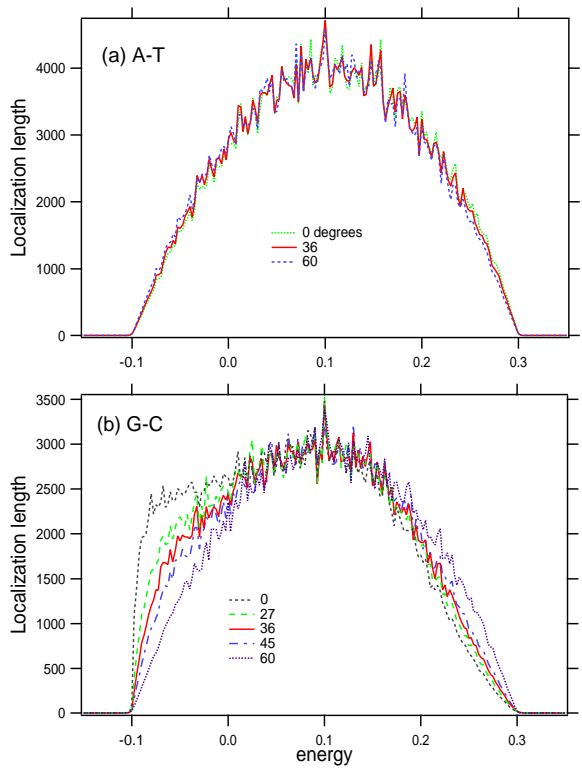}
 \caption{
The localization length as a function of
energy for several values of the twist angle $\theta_0(=0^\circ,
27^\circ, 36^\circ,45^\circ, 60^\circ)$
 in the (a) poly(dA)-poly(dT) DNA polymers,
(b)  poly(dG)-poly(dC)  DNA polymers and (c)  mixed DNA polymers.
The system size is
$N=2^{22}$ and $W=0.1$. }
\end{figure}

\subsection{System size effect on the resonance}
Figure 5 shows the system size dependence of the localization length
in the poly(dG)-poly(dC) polymer and the poly(dA)-poly(dT) polymer.
 Although the states for most of energies are exponentially localized, and 
have well-pronounced resonant maxima at discrete points correspond to the eigenenergies of
each system when the system size is relatively small ($N=2^{16}$). 
  Electronic states whose energy is close to resonance peaks have a tendency to be extended in the 
sense that their localization length is larger than the system size.
For the larger system size the energy dependence
of the localization length converges.

In our definition of the localization length, when the system size
$N$ is smaller than the localization length, the sharp energy
dependence of the resonance peaks appears, and the structure of
the resonance depends on the randomness in $r_{n}$ of each sample. 
Note that the resonance structure can be washed out if
we take the ensemble average $<\xi>$ with regard to
different configurations of $r_{n}$ even for $W=0.1$.

As a result we can say the transfer and transport of the electron sharply depend on 
the Fermi energy of the electroades as the system size is relatively small. 
The extended states caused by the resonance can 
contribute to the transfer of the electron.

\begin{figure}[!h]
\includegraphics[scale=.5]{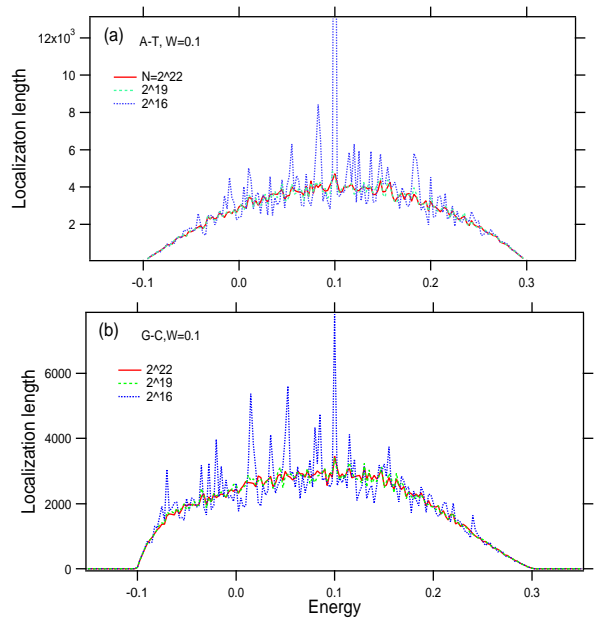}
 \caption{
The localization length as a function of
the energy for several system sizes
 $N(=2^{16}, 2^{19},2^{22})$ in the (a) poly(dA)-poly(dT) DNA
polymers, (b)  poly(dG)-poly(dC)  DNA polymers and (c)  mixed DNA
polymers. $W=0.1$ and $\theta_0=36^\circ$. }
\end{figure}

\section{Summary and Discussion}

We have numerically investigated localization properties of
electronic states in an adiabatic polaron model of poly(dG)-poly(dC) and poly(dA)-poly(dT)
DNA polymers with realistic parameters obtained using
semi-empirical  quantum-chemical calculations.

Now, we are in a position to compare the localization properties
of the poly(dG)-poly(dC), poly(dA)-poly(dT) DNA polymers and the
mixed case. Figure 6(a) and (b) show the localization length and
Lyapunov exponent (inverse of the localization length) in the
three types of the polymers with $W=0.1$. In the low energy
region, the localization length in the poly(dG)-poly(dC) DNA
polymer is larger than that in the poly(dA)-poly(dT) DNA polymer.
On the other hand, it is known from the experiments that
poly(dG)-poly(dC) oligomer is a better semiconductor than
poly(dA)-poly(dT) oligomer \cite{yoo01}. The difference is caused
by the dynamical effect which is neglected in this paper and the
system size effect. Indeed, the localization length of the DNA
polymers is larger than $\xi>2000[bp]$ in almost all the energy band  for all models, that is
much larger than the system size of the oligomer used in the
experiments. As seen in Fig.5 in the smaller system size the
resonance peaks become complex.

\begin{figure}[!h]
\includegraphics[scale=.55]{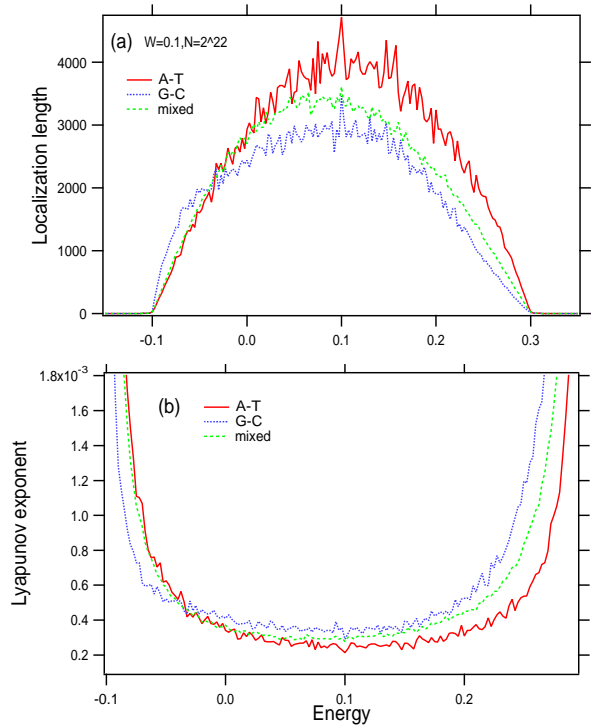}
 \caption{
Comparison (a)localization length and (b)Lyapunov exponent in
poly(dG)-poly(dC), poly(dA)-poly(dT) and the mixed DNA polymers.
$W=0.1$, $\theta_0=36^\circ$ and $N=2^{22}$.
}
\end{figure}

\begin{acknowledgments}
One of the authors (H.Y.)  would like to thank Dr. Kazumoto Iguchi
for his interest in this work.
\end{acknowledgments}


\end{document}